\documentclass{aastex}          
\usepackage{spr-astr-addons}    

\usepackage{hyperref}

\begin{document}
%
\title{Inhomogeneous imperfect fluid inflation}

\shorttitle{Inhomogeneous imperfect fluid inflation}
\shortauthors{E. Elizalde \& L. G. T. Silva}

\author{E. Elizalde\altaffilmark{1}}
\email{elizalde@ieec.uab.es}  
\and 
\author{Luis G. T. Silva\altaffilmark{2}}
\email{luisgustavot.silva@gmail.com}

\altaffiltext{1}{ICE-CSIC and IEEC, UAB Campus, C/ Can Magrans s/n, 08193 Bellaterra (Barcelona) Spain}
\altaffiltext{2}{Departamento de F\'{\i}sica, Universidade Estadual de Londrina, Rodovia Celso Garcia Cid, km 380, 
86057-970 Londrina - Paran\'{a}, Brazil}

\begin{abstract}
A generalized equation of state corresponding to a model that includes a Chaplygin gas and a viscous
term is investigated, in the context of the reconstruction program in scalar field cosmology.
The corresponding inflationary model parameters can be conveniently adjusted in order to reproduce the
most recent PLANCK data. The influence of the Chaplygin gas term contribution, in relation with
previous models, is discussed. Exit from inflation is shown to occur quite naturally in the new model.
\end{abstract}

\keywords{Inflation -- Dark energy -- Fluid models}

\section{Introduction} \label{s:introduction}
The analysis of the data obtained from astronomical observations has shown that we live in a spatially
flat universe in accelerated expansion \citep{ref:Ade:2015xua,ref:Ade:2015lrj}.
This acceleration could be produced by an hypothetical fluid with negative pressure, called dark
energy, which would approximately represent 70 percent of all energy contained in the universe,
according to the Standard Cosmological Model. The other also unobservable part of the universe is
considered by many to be constituted by weakly interacting massive particles (also considered sometimes
to form a fluid), and is termed as dark matter. This leaves only a mere 5 percent of the total energy
of the universe to ordinary matter.

The scientific community has been trying to explain the cosmic expansion of the universe using
different approaches but always taking into consideration the above percentage distribution into dark
matter and energy, and ordinary matter \citep{ref:Bamba:2012cp}. Different models have been considered,
as modified gravity \citep{ref:Myrzakulov:2013hca, ref:Nojiri:2006ri, ref:Nojiri:2010wj} where the cosmic
acceleration is provided without introducing exotic fluids. If one adopts a most natural and minimal
view, that fully keeps without modification the Einstein theory of gravity, two types of models have
been proposed: use of scalar fields and unified dark energy (UDE hereafter).
Scalar field models are thus called because they introduce a field, commonly called quintessence, which
is minimally couple with gravity and is self-interacting with the potential $V(\phi)$
\citep{ref:Peebles:2002gy}.
In the case of UDE models, it is assumed that dark energy and dark matter as different aspects of one
and the same, single fluid. In both context, scalar field and UDE approaches, two types of fluids have
lately received attention, namely the Chaplygin gas \citep{ref:Bento:2002ps, ref:Bilic:2001cg, ref:Gorini:2004by, 
ref:Xu:2012qx}
and a viscous fluid \citep{ref:Bamba:2015sxa, ref:Brevik:2015cya, ref:Capozziello:2005pa, ref:Haro:2015ljc,
ref:Myrzakulov:2014kta, ref:Nojiri:2005sr}). 

In this paper we adopt the method recently proposed by \cite{ref:Bamba:2014daa} to get a fluid
representation of the two main observables corresponding to inflationary models, namely the spectral
index $n_s$ of curvature fluctuations, and the tensor-to-scalar ratio $r$ of density fluctuations.
We propose a model containing a generalized viscous fluid with a Chaplygin gas term, and obtain the
allowed range of values of the model parameters in order to reproduce the most recent PLANCK results
\citep{ref:Ade:2015xua, ref:Ade:2015lrj}. It was shown by \cite{ref:Saadat:2013ava} that this type of fluid, with
some prearranged parameters, is a good model in order to reproduce the cosmological expansion. We go on
to analyze here the most general case.

The paper is organized as follows. In Sec.~\ref{s:description} the reconstruction program in scalar field
theory is briefly reviewed, and also the fluid description for the universe in accelerated expansion,
in a Friedmann-Lema\^{\i}tre-Robertson-Walker background. Following that, we recover the expressions for the
slow-roll parameters and for the observables of the inflationary model in the representation of fluid
models. In Sec.~\ref{s:fluid_model} we discuss our original model for a generalized viscous fluid with a
Chaplygin gas term. We explicitly show how the parameters corresponding to our model can be quite
naturally adjusted so as to reproduce the most recent and precise observational data. In Sec.~\ref{s:exit} we 
check the exit from inflation and the last section is devoted to conclusions.

\section{Inflation in the fluid model representation} \label{s:description}
In this section we summarize the procedure described by \cite{ref:Bamba:2014daa, ref:Bamba:2014wda}, which
we will use in what follows. The method simply consists in  rewriting first the slow-roll parameters as
a function of the Hubble parameter and its derivatives with respect to the number of e-folds.
Then the description of a perfect fluid model is adopted in order to rewrite the Hubble parameter, so
that the observables of the corresponding inflationary model can be finally written in the
representation of the fluid model.

\subsection{Slow-roll parameters} \label{ss:s-r_p}
Consider the action corresponding to the scalar field $\phi$ with the Einstein-Hilbert term
\begin{equation} \label{eqn:action}
S = \int d^4x\sqrt{-g}\left(\frac{R}{2\kappa^2}-\frac{1}{2}\partial_\mu\phi\partial^\mu\phi -
V(\phi)\right) \ ,
\end{equation}
where $g$ is the determinant of the metric $g_{\mu\nu}$ and $R$ the scalar curvature.
From action (\ref{eqn:action}) one gets the spectral index $n_s$ and the tensor-to-scalar ratio $r$, which
in the slow-roll regime are given by
\begin{equation} \label{eqn:oim}
n_s - 1\sim -6\epsilon + 2\eta \ , \quad r = 16\epsilon \ ,
\end{equation}
$\epsilon$ and $r$ being the slow-roll parameters, defined as
\begin{equation} \label{eqn:slow-roll}
\epsilon \equiv \frac{1}{2\kappa^2}\left(\frac{V'(\phi)}{V(\phi)}\right)^2 \ , \quad
\eta \equiv \frac{1}{\kappa^2}\frac{V''(\phi)}{V(\phi)} \ .
\end{equation}
Here and in what follows we will use the notation where the prime indicate derivative with respect to
the argument, for instance, $V'(\phi) \equiv \partial V(\phi)/\partial
\phi$.

As we are considering the flat Friedmann-Lema\^{\i}tre-Robertson-Walker (FLRW) universe, the metric is given by
\begin{equation}
ds^2 = - dt^2 + a^2(t)\sum_{i=1,2,3}(dx^i)^2 \ ,
\end{equation}
where $a(t)$ is the scale factor, which defines the Hubble parameter $H = \dot{a}/a$, the dot denoting
time derivative.
In this background, the gravitational field equations obtained from the action (\ref{eqn:action}) are
\begin{mathletters}
\begin{eqnarray} 
\frac{3}{\kappa^2}H^2 &=& \frac{1}{2}\dot{\phi}^2 + V(\phi) \ , \label{eqn:grav_f_eq_a} \\
-\frac{1}{\kappa^2}\left(H^2 + 2\dot{H}\right) &=& \frac{1}{2}\dot{\phi}^2 + V(\phi) \ . \label{eqn:grav_f_eq_b}
\end{eqnarray}
\end{mathletters}
Use of the already mentioned formulation by \cite{ref:Bamba:2014daa, ref:Bamba:2014wda, ref:Bamba:2015sxa}, yields
\begin{mathletters}
\begin{eqnarray}
\omega(\phi) &=& \left. - \frac{2}{\kappa^2}\frac{H'(N)}{H(N)}\right|_{N=\varphi} \ , \label{eqn:nfeqs_a} \\
V(\phi) &=& \left. \frac{1}{\kappa^2}H(N)^2\left(3 + \frac{H'(N)}{H(N)}\right)\right|_{N=\varphi} \ . \label{eqn:nfeqs_b}
\end{eqnarray}
\end{mathletters}
These equations follow from the solution of the gravitational field equations (\ref{eqn:grav_f_eq_a}) and 
(\ref{eqn:grav_f_eq_b}), wherethe scalar field  $\phi$ is replaced by a new scalar field
$\varphi$, $\phi = \phi(\varphi)$, and the positive quantity $\omega(\varphi) \equiv (d\phi/d\varphi)^2 > 0$ is 
introduced. Moreover $\varphi$ is identify with the number of e-folds
$N (\equiv	\ln(a_f/a_i) = \int_{t_i}^{t_f} Hdt)$, as a solution of the equation of motion for $\phi$ or
$\varphi$. With the quantities in Eqs.~(\ref{eqn:nfeqs_a}) and (\ref{eqn:nfeqs_b}), we are now able to express the 
slow-roll parameters in terms of $H(N)$ and its derivatives \citep[see][]{ref:Bamba:2014daa}.

\subsection{Fluid model description} \label{ss:fluid_model}
Continuing the procedure, we use the equation of state (EoS) of a fluid, as commonly used in fluid
models
\begin{equation} \label{eqn:eos}
P(N) = -\rho(N) + f(\rho) \ ,
\end{equation}
where $f(\rho)$ is an arbitrary function of the energy density $\rho(N)$, and $P(N)$ is the pressure of the fluid. 
The energy density and the pressure are given by \citep{ref:Bamba:2015sxa, ref:Bamba:2014wda}
\begin{eqnarray}
\rho(N) &=& \frac{3}{\kappa^2}H(N)^2 \ , \label{eqn:grav_eq1} \\
P(N) &=& -\frac{1}{\kappa^2}\left(2H(N)H'(N)+3H(N)^2\right) \ , \label{eqn:grav_eq2}
\end{eqnarray}
in the FLRW background.

Taking advantage of Eq.~(\ref{eqn:eos}), the conservation law can be rewritten as $\rho'(N)+3f(\rho)=0$, and
combining it with Eq.~(\ref{eqn:grav_eq2}), they yield
\begin{eqnarray} \label{eqn:H_f}
\frac{2}{\kappa^2}H(N)^2\left[\left(\frac{H'(N)}{H(N)}\right)^2 + \frac{H''(N)}{H(N)}\right] =
3f'(\rho)f(\rho) \ .
\end{eqnarray}
This equation allows us to finally express the slow-roll parameters (Eq.~(\ref{eqn:slow-roll})) in terms of
$\rho(N)$, $f(\rho)$ and corresponding derivatives, as can be seen in the appendix of \cite{ref:Bamba:2015sxa}
\begin{eqnarray}
\epsilon &=& \frac{3}{2}\frac{f(\rho)}{\rho(N)}\left(\frac{f'(\rho) - 2}{2 - f(\rho)/\rho(N)}\right)^2 \ , \label{eqn:epsilon} \\
\eta &=& \frac{3}{2 - f(\rho)/\rho(N)}\left[
\left(\frac{f(\rho)}{\rho(N)} + f'(\rho)\right)\right. \nonumber \\
&&\times\left.\left(1 - \frac{1}{2}f'(\rho)\right) -f(\rho)f''(\rho)\right] \ . \label{eqn:eta}
\end{eqnarray}
Consequently, in the same way as by \cite{ref:Bamba:2014wda}, we can express the observables of the
inflationary models (Eq.~(\ref{eqn:oim})) in the new representation, as
\begin{eqnarray}
n_s &\sim& 1 - 9\frac{f(\rho)}{\rho(N)}\left(\frac{2 - f'(\rho)}{2 - f(\rho)/\rho(N)}\right)^2 \nonumber \\
&&+ \frac{6}{2 - f(\rho)/\rho(N)}\left[\left(\frac{f(\rho)}{\rho(N)} + f'(\rho)\right)\right. \nonumber \\
&&\times \left.\left(1 - \frac{1}{2}f'(\rho)\right) - f(\rho)f''(\rho)\right] \ , \label{eqn:ns} \\
r &=& 24\frac{f(\rho)}{\rho(N)}\left(\frac{f'(\rho) - 2}{2 - f(\rho)/\rho(N)}\right)^2 \ . \label{eqn:r}
\end{eqnarray}
Concerning to this work, the advantage to use the approach described above is that we do not need integrate the
function $f(\rho)$, or terms like $1/f(\rho)$, which in our case can not be integrate maintaining the generality 
of the parameters.

\section{Generalized Chaplygin gas with viscosity} \label{s:fluid_model}
We here introduce our fluid model, with a generalized EoS with viscosity and a Chaplygin gas term,
namely
\begin{equation}
P = -\rho + A\rho^\alpha - B\rho^{-\beta} - \zeta(H) \ , \quad \zeta(H) = \bar{\zeta}H^\gamma \ .
\end{equation}
The positive constants $A$, $B$, $\bar{\zeta}$, $\alpha$, $\beta$ and $\gamma$ are here the
inflationary model parameters that we need to adjust in order to reproduce PLANCK's observational data 
\citep{ref:Ade:2015xua, ref:Ade:2015lrj}.
By comparing with Eq.~(\ref{eqn:eos}) and using Eq.~(\ref{eqn:grav_eq2}), we have
\begin{equation} \label{eqn:f_r}
f(\rho) = A\rho^\alpha -B\rho^{-\beta} - \tilde{\zeta}\rho^{\gamma/2} \ , \quad \tilde{\zeta} \equiv
\bar{\zeta}\left(\frac{\kappa}{\sqrt{3}}\right)^{\gamma} \ .
\end{equation}
The signs in Eq.~(\ref{eqn:f_r}) were set so that the energy density exponentially grows at early times and
becomes constant as time increases.
In the configuration $f(\rho) = -A\rho^\alpha - B\rho^{-\beta} +
\tilde{\zeta}\rho^{\gamma/2}$ we can find the same kind of behavior, but here the corresponding energy
density there diverges faster, for most possible values of the parameters.
Note that, if we set $\alpha = \gamma = 1, \beta = 0.5$ and $\bar{\zeta} = 3\zeta$ the model analyzed
by \cite{ref:Saadat:2013iu} is reproduced.
Further, the Chaplygin gas is dominant when $\rho$ is small, namely
\begin{equation} \label{eqn:CG}
f(\rho \ll 1) \approx -B\rho^{-\beta} \ .
\end{equation}
In the regime where $\rho$ is large we also obtain the same kind of EoS, assuming \citep{ref:Bamba:2015sxa}
$f(\rho) = A\rho^\alpha - \tilde{\zeta}\rho^{\gamma/2}$.
In this regime, different terms become dominant depending on whether $\alpha$ is larger or smaller than 
$\gamma/2$, respectively; in fact
\begin{eqnarray}
f(\rho \gg 1) &\approx& A\rho^\alpha \ , \ \text{for} \ \alpha > \gamma/2 \ , \label{eqn:hin} \\
f(\rho \gg 1) &\approx& -\tilde{\zeta}\rho^{\gamma/2} \ , \ \text{for} \ \alpha < \gamma/2 \ ,
\label{eqn:viscosity}
\end{eqnarray}
since we are assuming that $A$ and $\tilde{\zeta}$ are of the same order of magnitude.
The equations above (Eqs.~(\ref{eqn:CG}) to (\ref{eqn:viscosity})) tell us about the phantom barrier 
\citep{ref:Caldwell:2003vq}, where the sum of the pressure and the energy density (Eq.~(\ref{eqn:eos})) is no 
more negative and starts to be positive (that is $f(\rho) > 0$).

This barrier will be crossed if $\alpha > \gamma/2$, or for $\alpha < \gamma/2$ with the condition that $A$ is sufficiently larger than $\tilde{\zeta}$.
Because of the number of parameters that we have in Eq.~(\ref{eqn:f_r}), it is quite difficult to constraint
them properly by the observational data. What we will do in the next section is to admit a minimum preset and 
analyze how the slow-roll parameters and the observables behave when the rest of the parameters are
less than, larger or equal to the parameters set out initially. Models of the types below have been considered in \cite{ref:1986PhLB..180..335B, ref:1988NuPhB.310..743B, ref:1990PhLB..235...40B, ref:2004CQGra..21.5619B}.

\subsection{Constraining the model} \label{ss:constraining}
\begin{figure}[ht!]
 \centering
 \includegraphics[width=.7\columnwidth]{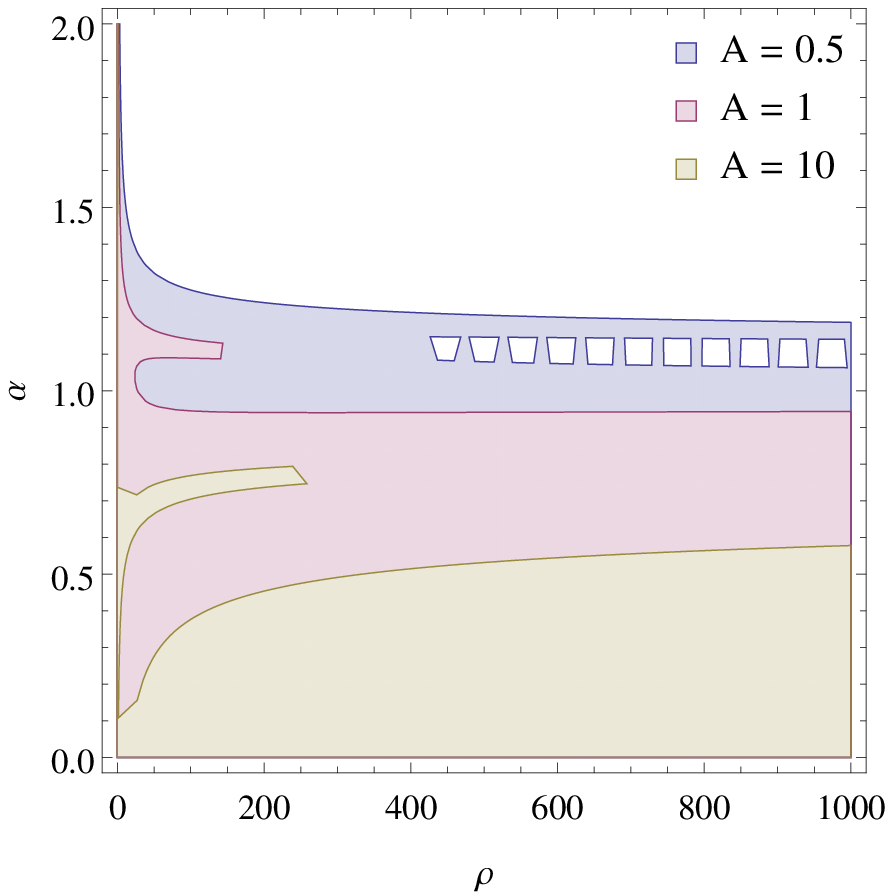}
 \caption{The region where $\epsilon < 1$ with variation on $\alpha$ with $B = \tilde{\zeta} = 1$, $\beta = 1/2$ and $\gamma = 1$.} 
 \label{fig:e_rxa1}
 \includegraphics[width=.7\columnwidth]{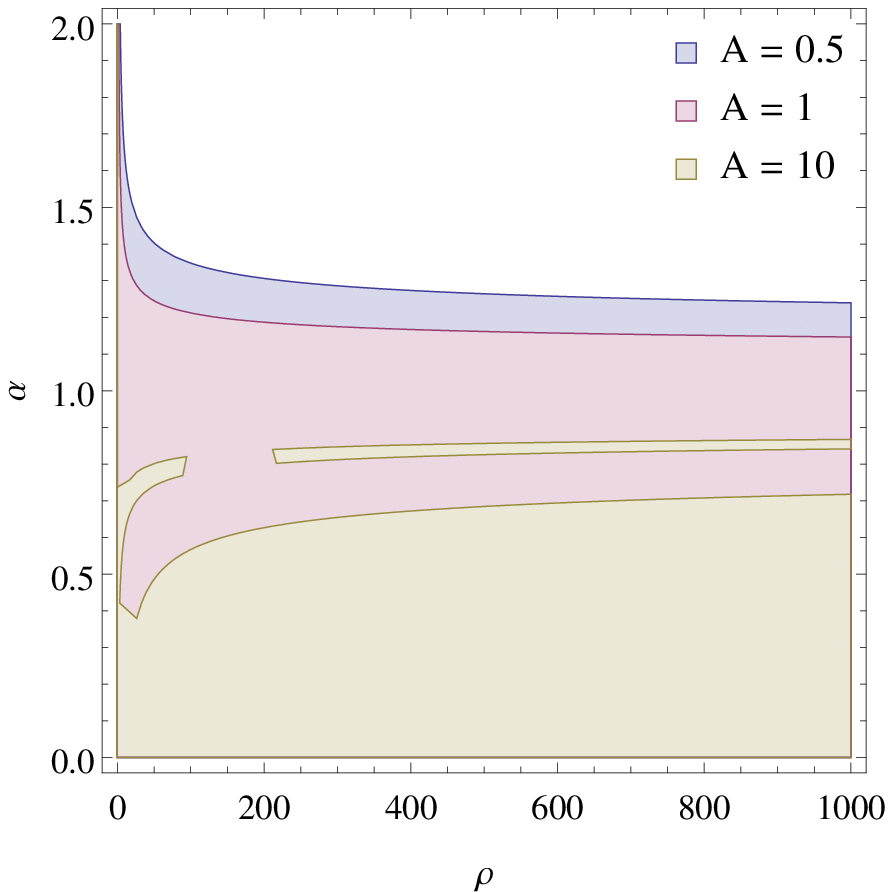}
 \caption{The region where $\epsilon < 1$ with variation on $\alpha$ with $B = \tilde{\zeta} = 1$, $\beta = 1/2$ and $\gamma = 2$.} 
 \label{fig:e_rxa2}
 \includegraphics[width=.7\columnwidth]{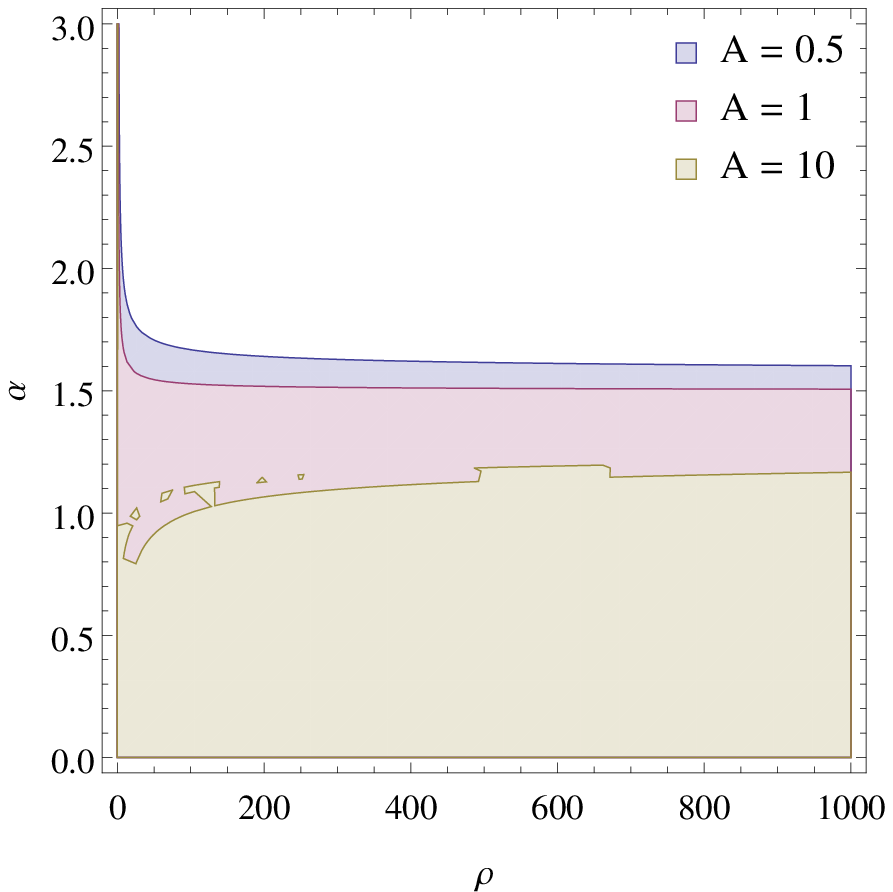}
 \caption{The region where $\epsilon < 1$ with variation on $\alpha$ with $B = \tilde{\zeta} = 1$, $\beta = 1/2$ and $\gamma = 3$.} 
 \label{fig:e_rxa3}
 \end{figure}

\begin{figure}[ht!]
 \centering
 \includegraphics[width=.7\columnwidth]{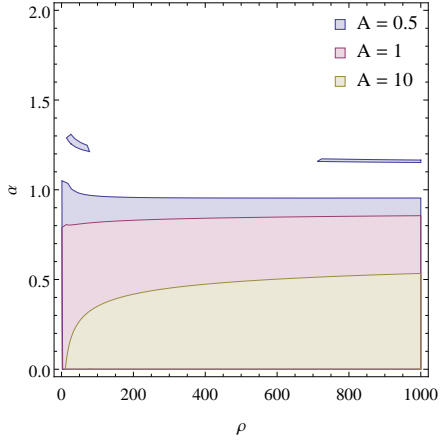}
 \caption{The region where $|\eta| < 1$ with variation on $\alpha$ with $B = \tilde{\zeta} = 1$, $\beta = 3/4$ and $\gamma = 1$.} 
 \label{fig:n_rxa}
 \end{figure}

\begin{figure}[ht!]
 \centering
 \includegraphics[width=.7\columnwidth]{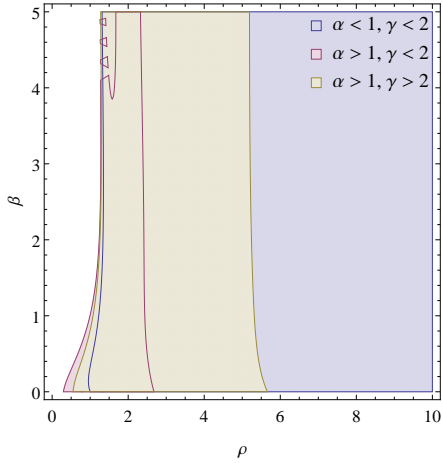}
 \caption{The region where $|\eta| < 1$ with variation on $\beta$ with $A = B = \tilde{\zeta} = 1$. The three cases are $\alpha = 0.75, \ \gamma = 1.5; \alpha = 1.5, \ \gamma = 1.5$ and $\alpha = 1.5, \ \gamma = 2.5$ respectively.} 
 \label{fig:n_rxb}
 \end{figure}

\begin{figure}[ht!]
 \centering
 \includegraphics[width=.7\columnwidth]{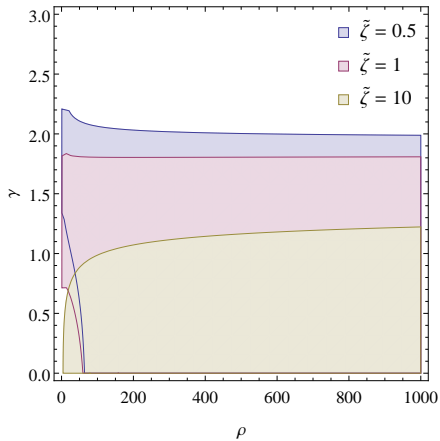}
 \caption{The region where $|\eta| < 1$ with variation on $\gamma$ with $A = B = 1$, $\alpha = 3/4$ and $\beta = 1/2$ and.} 
 \label{fig:n_rxg}
 \end{figure}

It is well know that, in order for inflation to occur the slow-roll parameters (Eq.~(\ref{eqn:slow-roll})) must 
satisfy the following constraints: $\epsilon, |\eta| < 1$ \citep{ref:Baumann:2009ds}.
If we want to keep the arbitrariness of the inflationary  parameters of our model, we are not able to
solve the Friedmann equations analytically. Thus we adopt the alternative procedure to apply the EoS
(Eq.~(\ref{eqn:f_r})) on Eqs.~(\ref{eqn:epsilon}) and (\ref{eqn:eta}) and then check numerically which is the 
allowed range of the
parameters in order to obey such constraints (Figs.~\ref{fig:e_rxa1} to \ref{fig:n_rxg}).

Since the second and the third term of Eq.~(\ref{eqn:f_r}) are negative, they only contribute  to make more 
negative the value of the slow-roll parameter $\epsilon$, because $f(\rho)/\rho$ is what determines the sign in 
Eq.~(\ref{eqn:epsilon}).
Then, we have only to worry about the first term.
In Figs.~\ref{fig:e_rxa1} to \ref{fig:e_rxa3} we see that the contribution of the Chaplygin gas term allows 
the range in $\alpha$ to increase for small energy density, whereas $A \leq \tilde{\zeta}$.
For large $\rho$, the $\alpha$ values begin to get more limited, especially if the coefficient $A$ is
big, as compared with the other coefficients.
In fact, when $\rho \gg 1$ the term in brackets in Eq.~(\ref{eqn:epsilon}) tends to one, and we have
\begin{equation} \label{eqn:epsilon_asy}
\epsilon \approx \frac{3}{2}\left(A\rho^{-1 + \alpha} - \tilde{\zeta}\rho^{-1 + \gamma/2}\right) \ .
\end{equation}
If $\gamma < 2$, then $\epsilon \approx (3/2)A\rho^{-1+\alpha}$, and in the limit case, when $\rho
\rightarrow \infty$ it is required that $\alpha \rightarrow 1$, in order to satisfy the slow-roll
condition (Fig.~\ref{fig:e_rxa1}).
In this case the upper bound in $\alpha$ decreases, as the coefficient $A$ increases, no matter the
value of $\tilde{\zeta}$.
If $\gamma = 2$ the slow-roll parameter becomes $\epsilon \approx (3/2)(A\rho^{-1+\alpha} -
\tilde{\zeta})$ and the same limit $\alpha \rightarrow 1$ is required, but in this case with a little
delay, because of $\tilde{\zeta}$ (Fig.~\ref{fig:e_rxa2}).
For $\gamma > 2$ we have  Eq.~(\ref{eqn:epsilon_asy}) and $\alpha \rightarrow \gamma/2$, when $\rho
\rightarrow \infty$ (Fig.~\ref{fig:e_rxa3}).
For the last two cases the upper bound in $\alpha$ decrease as $A$ becomes larger than
$\tilde{\zeta}$. 

In Figs.~\ref{fig:n_rxa} to \ref{fig:n_rxg} one can spot the region where the condition $|\eta| < 1$ is 
satisfied.
Because of the number of terms in Eq.~(\ref{eqn:eta}), various combinations of the parameters will fulfill the
constraint. Basically, it is easy to understand that, since we have $\alpha \lesssim 1$ and $\gamma \lesssim 2$ 
the slow-roll condition will be satisfied when the energy density becomes large, with a big range for $\beta$ 
(Fig.~\ref{fig:n_rxb}).
We can see this upper bound in $\alpha$ and $\gamma$ in Figs.~\ref{fig:n_rxa} and \ref{fig:n_rxg}.
Since all the coefficients are of the same order, the Chaplygin gas does no affect the region $\rho \gg 1$ and the 
value of $\beta$ does not change the behavior in Figs.~\ref{fig:n_rxa} and \ref{fig:n_rxg} in this regime.
Furthermore, the increase in the coefficient $A$ and $\tilde{\zeta}$ lowers the upper limit of $\alpha$ and $\gamma$ respectively.


\begin{figure}[ht!]
 \centering
 \includegraphics[width=.7\columnwidth]{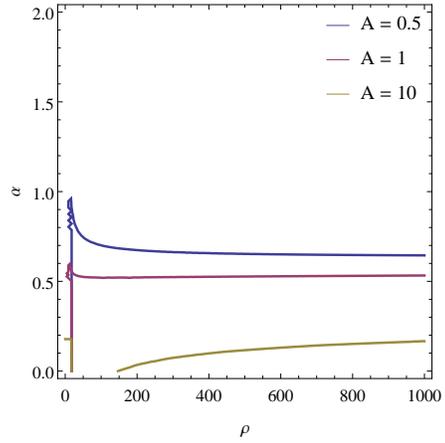}
 \caption{The curves $n_s = 0.968$ with variation on $\alpha$ with $B = \tilde{\zeta} = 1$, $\beta = 1/2$ and $\gamma = 1$.} 
 \label{fig:ns_rxa}
 \end{figure}

\begin{figure}[ht!]
 \centering
 \includegraphics[width=.7\columnwidth]{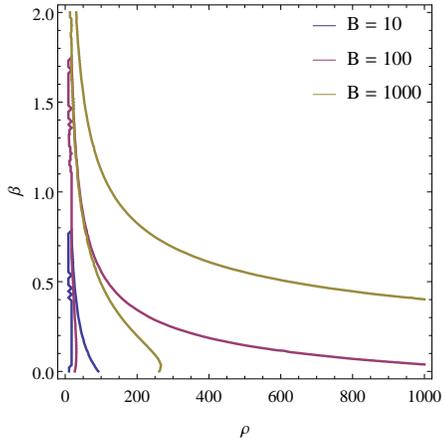}
 \caption{The curves $n_s = 0.968$ with variation on $\beta$ with $A = \tilde{\zeta} = 1$, $\alpha = 3/4$ and $\gamma = 1$.} 
 \label{fig:ns_rxb}
 \end{figure}

\begin{figure}[ht!]
 \centering
 \includegraphics[width=.7\columnwidth]{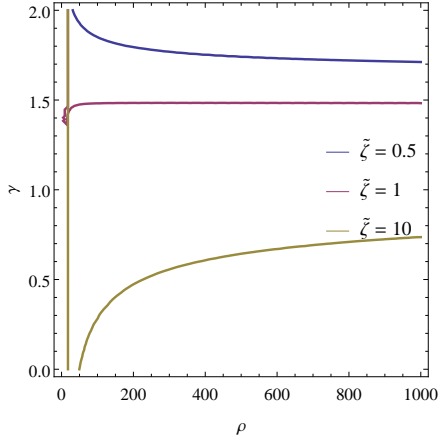}
 \caption{The curves $n_s = 0.968$ with variation on $\gamma$ with $A = B = 1$, $\alpha = 3/4$ and $\beta = 1/2$.} 
 \label{fig:ns_rxg}
 \end{figure}

\begin{figure}[ht!]
 \centering
 \includegraphics[width=.7\columnwidth]{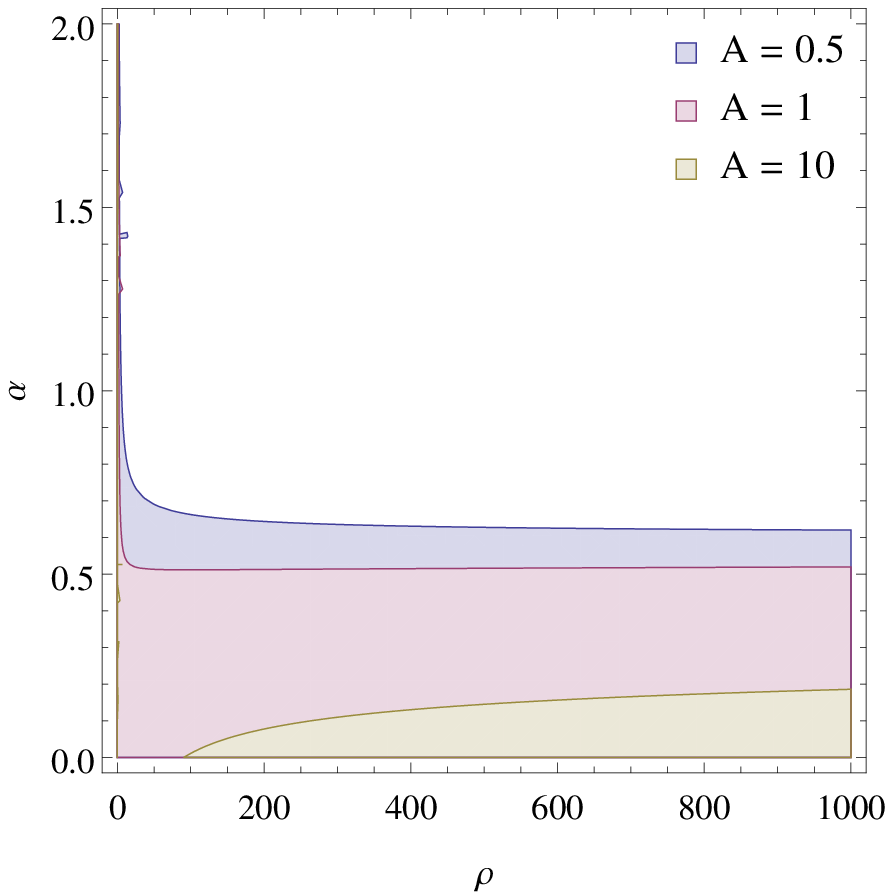}
 \caption{The region where $r < 0.11$ with variation on $\alpha$ with $B = \tilde{\zeta} = 1$, $\beta = 1/2$ and $\gamma = 1$.} 
 \label{fig:r_rxa}
 \end{figure}

\begin{figure}[htb!]
 \centering
 \includegraphics[width=.7\columnwidth]{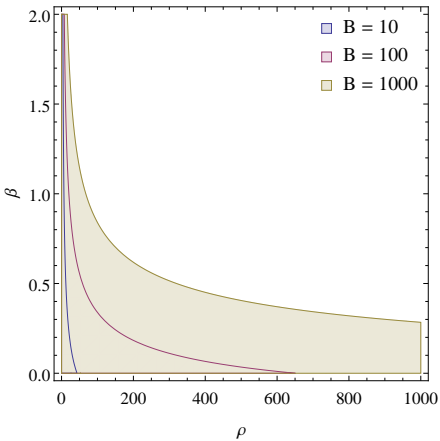}
 \caption{The region where $r < 0.11$ with variation on $\beta$ with $A = \tilde{\zeta} = 1$, $\alpha = 3/4$ and $\gamma = 1$.} 
 \label{fig:r_rxb}
 \end{figure}

\begin{figure}[htb!]
 \centering
 \includegraphics[width=.7\columnwidth]{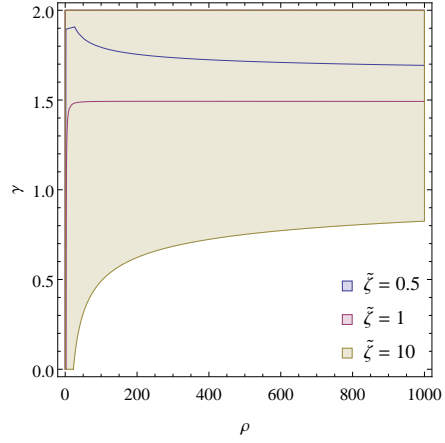}
 \caption{The region where $r < 0.11$ with variation on $\gamma$ with $A = B = 1$, $\alpha = 3/4$ and $\beta = 1/2$.} 
 \label{fig:r_rxg}
 \end{figure}

We now compare with the most recent observational results by PLANCK.
To do that, we have applied the EoS (Eq.~(\ref{eqn:f_r})) in Eqs.~(\ref{eqn:ns}) and (\ref{eqn:r}) and used the
PLANCK results, $n_s = 0.968 \pm 0.006 (68\% \ \text{CL})$ and $r < 0.11 (95\% \ \text{CL})$ 
\citep{ref:Ade:2015xua, ref:Ade:2015lrj}, to analyze the inflationary model parameters (Figs.~\ref{fig:ns_rxa} 
to (\ref{fig:r_rxg})).

Even if the condition on $n_s$ becomes more restrictive, the parameters $\alpha$, $\beta$ and $\gamma$
are subject to the coefficients $A$, $B$ and $\tilde{\zeta}$, and we cannot estimate  specific values
for them (Figs.~\ref{fig:ns_rxa} to \ref{fig:ns_rxg}).
In Figs.~\ref{fig:ns_rxb} and \ref{fig:r_rxb} we see that the Chaplygin gas term must have more influence than 
the other terms
($B \gg A,\tilde{\zeta}$), in order to contribute to the observational data when $\rho \gg 1$.
In this case, it seems reasonable to say that $0 < \beta \lesssim 0.5$. We will here maintain this
term, which was absent by \cite{ref:Bamba:2015sxa}, to see how it can possibly modified the results.
From Figs.~\ref{fig:ns_rxa}, \ref{fig:ns_rxg}, \ref{fig:r_rxa} and \ref{fig:r_rxg} we can check what was 
already discussed about the slow-roll
parameter $\epsilon$, that is $\alpha \rightarrow \gamma/2$ (or $\gamma \rightarrow 2\alpha$) when
$\rho \rightarrow \infty$.

Note that we have a lower bound on $\gamma$ (Fig.~\ref{fig:r_rxg}).
Thus, in order to satisfy the slow-roll condition and to reproduce the observational data we have two
scenarios. When $A > \tilde{\zeta}$ the safety range for $\gamma$ is bound to [2$\alpha$,2] where the borders are 
achieved when $\rho \rightarrow \infty$.
(In terms of $\alpha$ we can say that its range is bound to [0,$\gamma/2$]).
On the other hand, when $A \leqslant \tilde{\zeta}$ the lower bound on $\gamma$ is less than $2\alpha$ or, 
conversely, the upper bound $\alpha$ is larger than $\gamma/2$.
In this case, by Eq.~(\ref{eqn:hin}), the phantom barrier can be crossed and the range is set based on the
slow-roll condition shown in Figs.~\ref{fig:e_rxa1} to \ref{fig:n_rxg}, $0 < \alpha < 1$ and $0 < \gamma < 2$.
In this case the limit $\alpha \rightarrow \gamma/2$ is also reached when $\rho \rightarrow \infty$.
In fact, the grater the difference $A - \tilde{\zeta}$ grater $\rho$ needs to be for the limit is reached.

 

\section{Exit from inflation} \label{s:exit}
We need now check if, in our fluid model, the universe is actually able to exit from the inflationary stage and to 
continue its evolution to subsequent reheating regimes. Otherwise, the accelerate expansion phase would never 
stop. In other words, we need to analyze the instability of the corresponding de Sitter solution, 
$H = H_{\text{inf}}$, a positive constant.
This analysis will proceed in a way similar to that by \cite{ref:Bamba:2015sxa}, where the Hubble
parameter is written as follow
\begin{equation} \label{eqn:deS}
H = H_{\text{inf}} + H_{\text{inf}}\delta(t) \ ,
\end{equation}
$|\delta(t)| \ll 1$ being a small perturbation which we define as
\begin{equation} \label{eqn:delta}
\delta(t) \equiv e^{\lambda t} \ .
\end{equation}
Here, $\lambda$ is a constant that will give us information about the instability of the de Sitter
solution.
In order to proceed with the analysis, we rewrite Eq.~(\ref{eqn:H_f}) with respect to the cosmic time $t$ and
apply Eq.~(\ref{eqn:f_r}); this yields
\begin{align} \label{eqn:second_der_H}
\ddot{H} &- \frac{\kappa^4}{2}\left[
\alpha A^2\left(\frac{3}{\kappa^2}\right)^{2\alpha}H^{4\alpha-1}\right. \nonumber \\
&- \beta B^2\left(\frac{3}{\kappa^2}\right)^{-2\beta}H^{-4\beta-1}
+ \frac{\gamma}{2}\tilde{\zeta}^2\left(\frac{3}{\kappa^2}\right)^\gamma H^{2\gamma-1} \nonumber \\
&- (\alpha - \beta)AB\left(\frac{3}{\kappa^2}\right)^{\alpha-\beta}H^{2(\alpha-\beta)-1} \nonumber \\
&- \left(\alpha + \frac{\gamma}{2}\right)A\tilde{\zeta}\left(\frac{3}{\kappa^2}\right)^{\alpha + \gamma/2}
H^{2\alpha+\gamma-1} \nonumber \\
&- \left(\beta - \frac{\gamma}{2}\right)B\tilde{\zeta}\left.\left(\frac{3}{\kappa^2}\right)^{-\beta + \gamma/2}
H^{-2\beta+\gamma-1}\right]= 0 \ .
\end{align}
Substituting Eq.~(\ref{eqn:deS}) with Eq.~(\ref{eqn:delta}) into Eq.~(\ref{eqn:second_der_H}) and taking the first 
order in $\delta(t)$, we get
\begin{equation} \label{eqn:lambda}
\lambda^2 - \frac{1}{2}\frac{\kappa^4}{H_{\text{inf}}}Q = 0 \ ,
\end{equation}
where we have defined
\begin{align}
Q &\equiv \alpha(4\alpha - 1)A^2\left(\sqrt{3}\frac{H_{\text{inf}}}{\kappa}\right)^{4\alpha} \nonumber \\
&+ \beta(4\beta - 1)B^2\left(\sqrt{3}\frac{H_{\text{inf}}}{\kappa}\right)^{-4\beta} \nonumber \\
&+ \frac{\gamma}{2}(2\gamma -1)
\tilde{\zeta}^2\left(\sqrt{3}\frac{H_{\text{inf}}}{\kappa}\right)^{2\gamma} \nonumber \\
&- (\alpha - \beta)\left[2(\alpha - \beta) -
1\right]AB\left(\sqrt{3}\frac{H_{\text{inf}}}{\kappa}\right)^{2(\alpha-\beta)} \nonumber \\
&- \left(\alpha + \frac{\gamma}{2}\right)(2\alpha + \gamma -
1)A\tilde{\zeta}\left(\sqrt{3}\frac{H_{\text{inf}}}{\kappa}\right)^{2\alpha+\gamma} \nonumber \\
&- \left(\beta - \frac{\gamma}{2}\right)(-2\beta + \gamma -1) \nonumber \\
&\times B\tilde{\zeta}\left(\sqrt{3}\frac{H_{\text{inf}}}{\kappa}\right)^{-2\beta+\gamma} \ .
\end{align}
Note that, if we set $B = 0$ we obtain the same result as by \cite{ref:Bamba:2015sxa}, since this
coefficient determines the contribution of the Chaplygin gas.
Another interesting point to note is that if $\alpha = \beta = \gamma/2 = 1/4$ we obtain $Q = 0$,
independently of the values of the coefficients, and from Eq.~(\ref{eqn:lambda}) the de Sitter solution is 
directly recovered.
One can easily note that by Eq.~(\ref{eqn:f_r}) this case reproduces a expanded Chapligyn gas,
$f(\rho) = (A - \tilde{\zeta})\rho^{1/4} - B\rho^{-1/4}$.

The two solutions of Eq.~(\ref{eqn:lambda}) are
\begin{equation}
\lambda = \lambda_\pm \equiv \pm\frac{1}{\sqrt{2}}\frac{\kappa^2}{H_{\text{inf}}}\sqrt{Q} \ .
\end{equation}
We are looking for the positive solution $\lambda = \lambda_+ > 0$, which is obtained if $Q > 0$.
In this way, we see that, as the cosmic time grows, $\delta(t)$ becomes larger and the exit from
inflation will occur in a natural way.

\begin{figure}[ht!]
 \centering
 \includegraphics[width=.7\columnwidth]{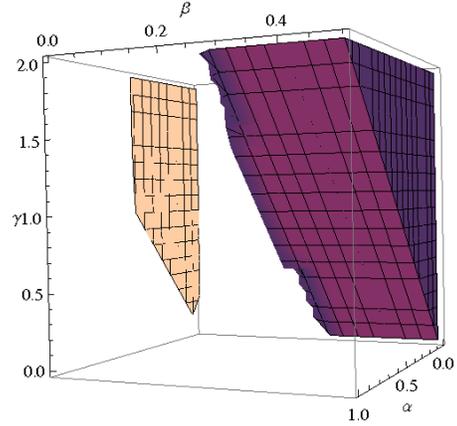}
 \caption{The region where $Q > 0$ for $A > \tilde{\zeta}$ scenario with $H_{\text{inf}}/\kappa = 0.1$, $B = 10^3$, $A = 10$ and $\tilde{\zeta} = 1$.} 
 \label{fig:Q_sce_1a}
 \end{figure}

\begin{figure}[ht!]
 \centering
 \includegraphics[width=.7\columnwidth]{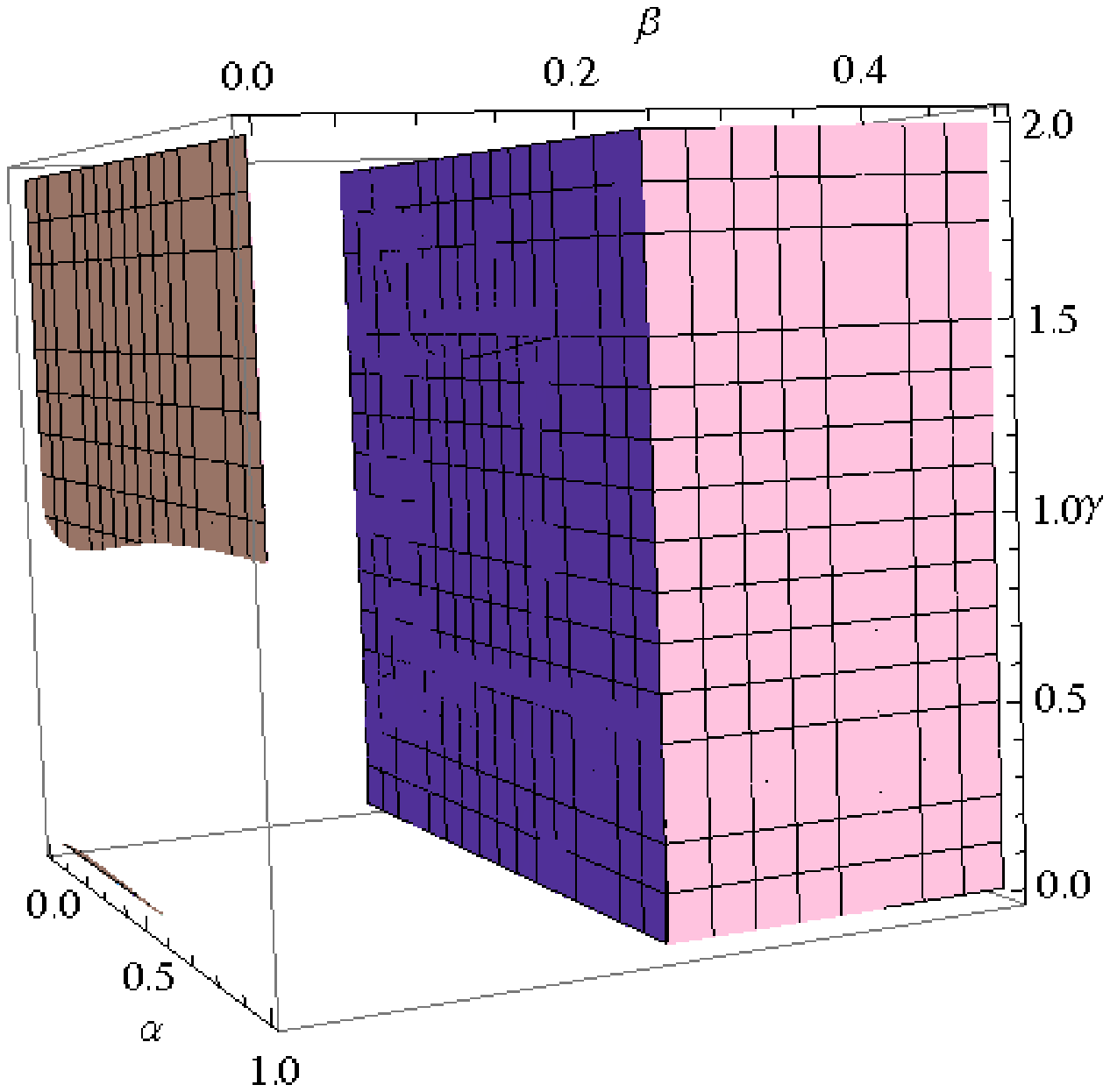}
 \caption{The region where $Q > 0$ for $A \leqslant \tilde{\zeta}$ scenario with $H_{\text{inf}}/\kappa = 0.1$, $B = 10^3$, $A = 1$ and $\tilde{\zeta} = 10$.} 
 \label{fig:Q_sce_2a}
 \end{figure}

In the last section (Sec.~\ref{ss:constraining}) we found two scenarios, one in which $A > \tilde{\zeta}$ and
other for $A \leqslant \tilde{\zeta}$.
If $H_{\text{inf}}/\kappa \ll 1$ we obviously see the influence of the Chapligyn gas parameter as shown in 
Figs.~\ref{fig:Q_sce_1a} and \ref{fig:Q_sce_2a}.
In both cases solutions for $\beta = 0$ and $0.25 < \beta < 0.5$ was found.
For the others parameters we have
\begin{eqnarray}
\text{Scenario 1} \ (A > \tilde{\zeta})	&:& \ 0 < \alpha < 1, \ 0 < \gamma < 2, \nonumber \\
										&&\text{with} \ \alpha < \gamma/2 \ ,
\end{eqnarray}
and
\begin{equation}
\text{Scenario 2} \ (A \leqslant \tilde{\zeta}): \ 0 < \alpha < 1, \ 0 < \gamma < 2 \ ,
\end{equation}


\begin{figure}[ht!]
 \centering
 \includegraphics[width=.7\columnwidth]{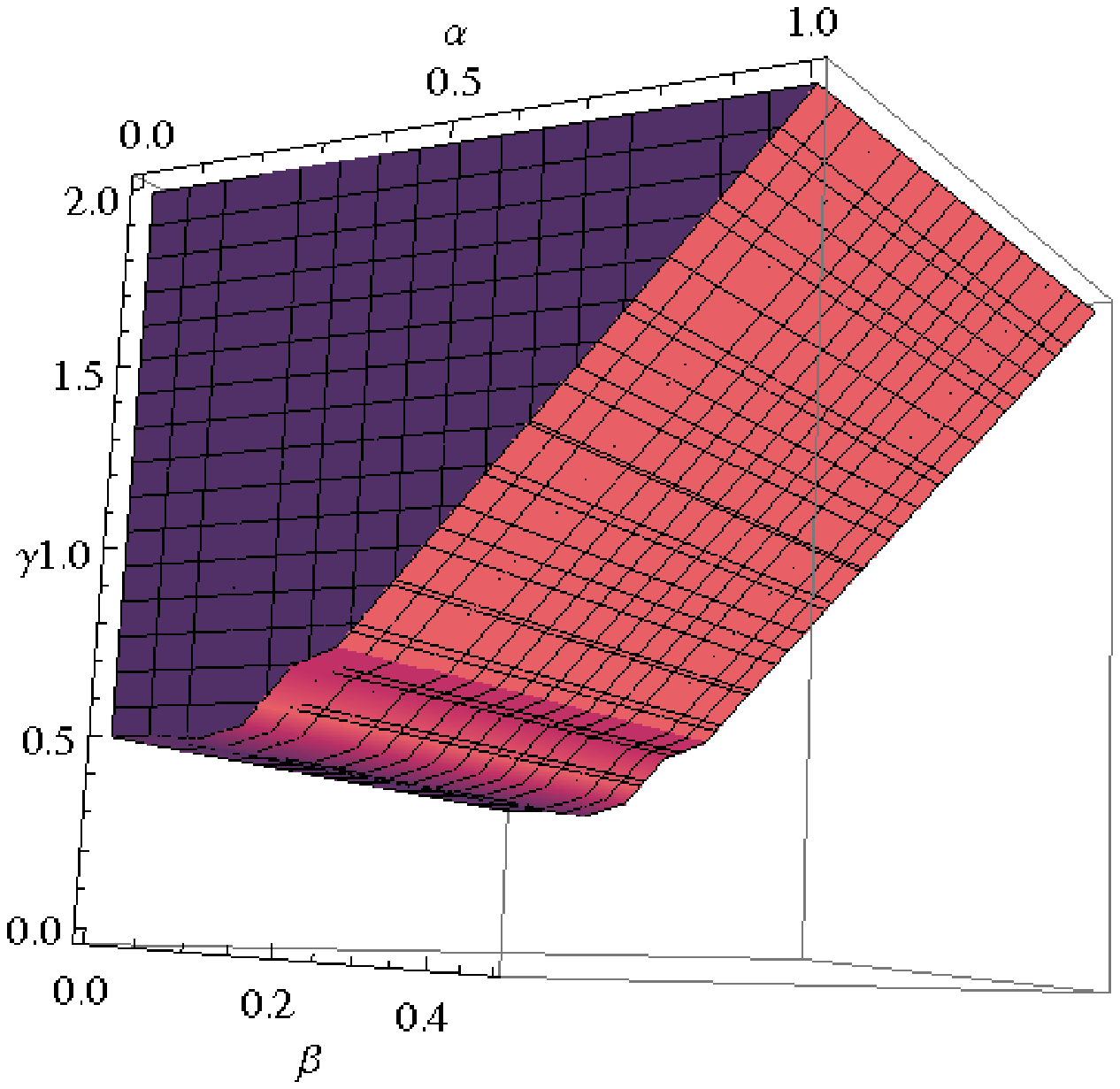}
 \caption{The region where $Q > 0$ for $A > \tilde{\zeta}$ scenario with $H_{\text{inf}}/\kappa = 10^{10}$, $B = 10^3$, $A = 10$ and $\tilde{\zeta} = 1$.} 
 \label{fig:Q_sce_1b}
 \includegraphics[width=.7\columnwidth]{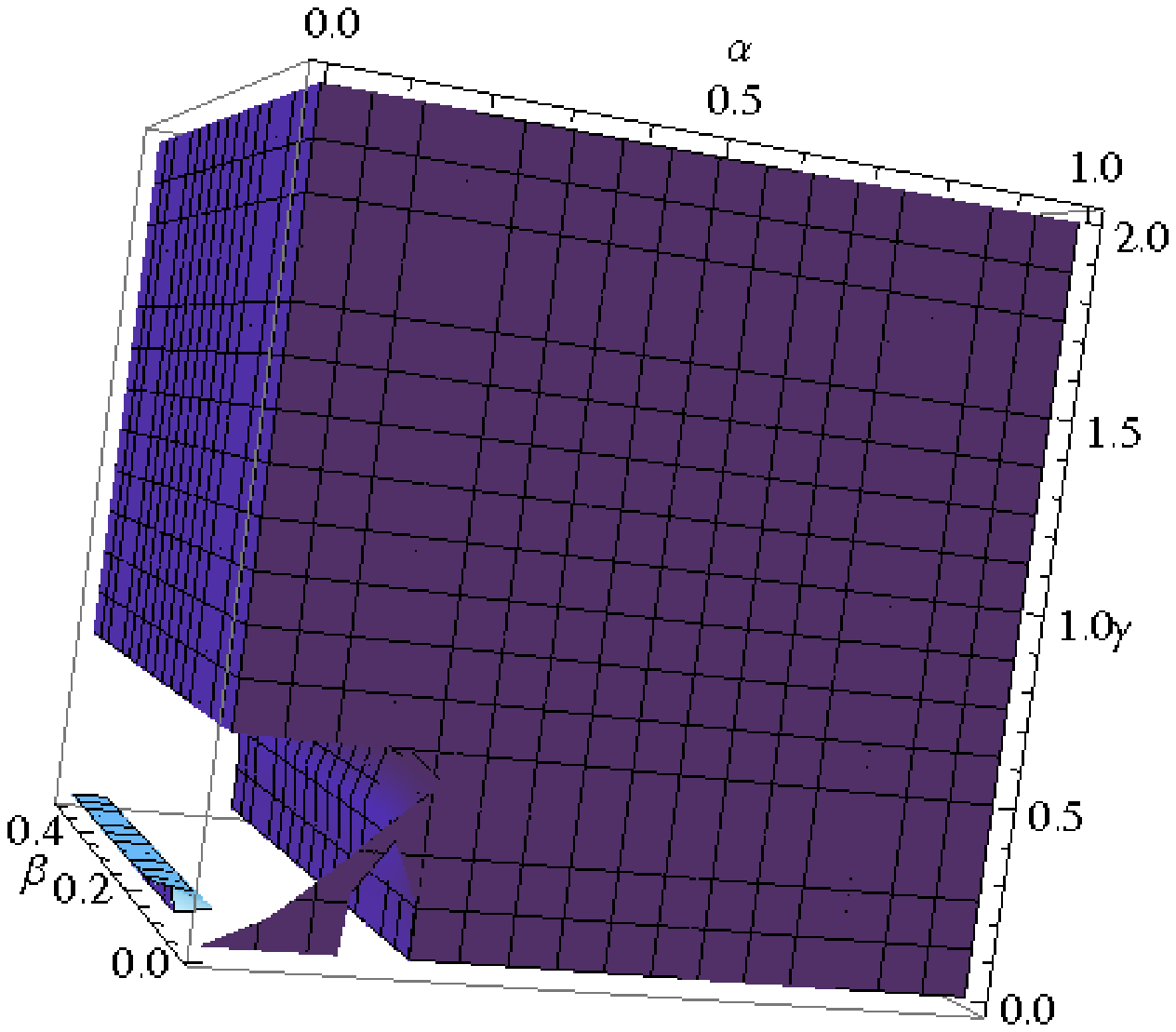}
 \caption{The region where $Q > 0$ for $A \leqslant \tilde{\zeta}$ scenario with $H_{\text{inf}}/\kappa = 10^{10}$, $B = 10^3$, $A = 1$ and $\tilde{\zeta} = 10$.} 
 \label{fig:Q_sce_2b}
 \end{figure}

For the case which $H_{\text{inf}} \gg 1$ we have no restriction on $\beta$ (unless 
$\alpha \approx \gamma \approx 0$ and/or $B \ll H_{\text{inf}}$) and the values of the other parameters
becomes important.
Summary we obtain, as shown in Figs.~\ref{fig:Q_sce_1b} and \ref{fig:Q_sce_2b},
\begin{eqnarray}
\text{Scenario 1} \ (A > \tilde{\zeta})	&:& \ 0 < \alpha < 1, \ 0.5 < \gamma < 2, \nonumber \\
										&&\text{with} \ \alpha < \gamma/2 \ ,
\end{eqnarray}
and
\begin{eqnarray}
\text{Scenario 2} \ (A \leqslant \tilde{\zeta}):
	\begin{cases}
	0 < \alpha < 1, \ 0.5 < \gamma < 2, \\
	\text{and} \\
	0.25 < \alpha < 1, \ 0 < \gamma < 2 \ .
	\end{cases}
\end{eqnarray}

\section{Conclusions} \label{s:conc}
In this work we have investigated an inhomogeneous imperfect fluid with a Chaplygin gas term and
viscosity, as a model in order to explain the cosmic accelerated expansion. We have managed to adjust the 
parameters of the model in order to reproduce PLANCK's most recent observational
results. To carry out this task we used the established procedure of reconstruction of a scalar
field theory of inflation, with the purpose to describe the slow-roll parameters and, consequently, the 
observables of the inflationary model, in terms of their representation as a fluid model, based on previous works 
of \cite{ref:Bamba:2014daa, ref:Bamba:2014wda, ref:Bamba:2015sxa}.

By comparing with PLANCK's data, we have determined that the effect of the Chaplygin gas has to be
definitely bigger than the contribution of the other terms, if one wants to see some trace of this term in the 
$\rho \gg 1$ regime. In this case, we have found the range $0 < \beta < 0.5$.
For the rest of the parameters we determined two different scenarios: (1) one with $A > \tilde{\zeta}$, where 
$0 < \alpha < \gamma/2$, $2\alpha < \gamma < 2$, and (2) another with $A \leqslant \tilde{\zeta}$ where 
$0 < \alpha < 1$, $0 < \gamma < 2$.
We have also verified that in our new model model the exit from inflation can occur quite naturally.
As a result of our analysis, one more small restriction was obtained for each scenario, namely 
$\gamma \gtrsim 0.5$ for case (1), and $\alpha \gtrsim 0.4$, $\gamma \gtrsim 0.6$ for case (2),
with $0 < \beta < 0.5$ for $H_{\text{inf}}/\kappa \gg 1$.

We should finally point out that using these methods the possibility of a viscous LR cosmology (as in, e.g., 
\cite{ref:Brevik:2011mm}) can be studied, too.


 \acknowledgments
EE has been supported in part by MINECO (Spain), project FIS2013-44881, by I-LINK1019 from CSIC, and by the CPAN 
Consolider Ingenio Project.
LGTS has been supported by a grant from CAPES Foundation through the PDSE program, process number: 
99999.003658/2015-05.

%


\end{document}